\documentclass[12pt]{article}
\usepackage{times}
\usepackage{amsmath}
\usepackage{amssymb}
\usepackage{graphicx}
\usepackage{fancyhdr}
\usepackage{euscript,fancyhdr,graphicx,cite,ifthen,makeidx,multicol,pifont,psboxit,pstricks}
\let\savehline\hline
\def\sphline{\noalign{\vskip3pt}\savehline\noalign{\vskip3pt}}
\newcommand{\balpha}{{\mbox {\boldmath $ \alpha $}}}
\newcommand{\bPsi}{{\mbox {\boldmath $ \Psi $}}}
\newcommand{\bS}{{\mbox {\boldmath $ S $}}}
\newcommand{\by}{{\mbox {\boldmath $ y $}}}

\newcommand{\bw}{{\mbox {\boldmath $ w $}}}
\newcommand{\bz}{{\mbox {\boldmath $ z $}}}
\newcommand{\E}{\mbox{E}}
\newcommand{\N}{N_0}
\newcommand{\beq}{\begin{equation}}
\newcommand{\eeq}{\end{equation}}
\date{{\sl \today}}

\def\insertfig#1#2#3#4#5{
\begin{figure}[#1]
\centering\includegraphics[width=#2\columnwidth,clip=]{#3.eps}
\caption{#4}\label{#5}\end{figure}}
\def\insertfig#1#2#3#4#5{
\begin{figure}[#1]
\centering\includegraphics[width=#2\columnwidth,clip=]{#3.eps}
\caption{#4}\label{#5}\end{figure}}

\newcommand{\threshold}{\mbox{$\begin{array}{c} \stackrel{\textstyle H_1}{\textstyle>}
\\[-4pt]
\stackrel{\textstyle <}{\textstyle H_0}
\end{array}$}}
\begin{document}
\title{Neighbor Discovery in Wireless Networks: \\[-2pt] A Multiuser-Detection Approach}
\author{
\begin{minipage}[t]{ 5cm}
\centering
$\;$ \\
{\large Daniele Angelosante\footnote{DAEIMI, Universit\`a di
Cassino, Italy. email: {\tt d.angelosante@unicas.it }}}
\end{minipage}
\begin{minipage}[t]{ 5cm}
\centering
$\;$ \\
{\large Ezio Biglieri\footnote{Departament de Tecnologia,
Universitat Pompeu Fabra, Barcelona, Spain. email: {\tt
ezio.biglieri@upf.edu}. This work was supported by the STREP
project No.\ IST-026905 (MASCOT) within the 6th framework program
of the European Commission, and by the Spanish Ministery of
Education and Science under Project TEC2006-01428/TCM.}}
\end{minipage}
\begin{minipage}[t]{5cm}
\centering
$\;$ \\
{\large Marco Lops\footnote{DAEIMI, Universit\`a di Cassino,
Italy. email: {\tt lops@unicas.it}}}
\end{minipage}
}
 \maketitle
\begin{abstract}
We examine the problem of determining which nodes are neighbors of
a given one in a wireless network. We consider an unsupervised
network operating on a frequency-flat Gaussian channel, where
$K+1$ nodes associate their identities  to nonorthogonal
signatures, transmitted at random times, synchronously, and
independently. A number of neighbor-discovery algorithms, based on
different optimization criteria, are introduced and analyzed.
Numerical results show how reduced-complexity algorithms can
achieve a satisfactory performance.
\end{abstract}
%
%\IEEEpeerreviewmaketitle
%
\section{Introduction}\label{SEC1}

Of late, wireless networks, and in particular sensor networks,
have been the object of a good deal of interest, also spurred by
the manifold applications they can be associated with (see, for
example, their applications to classification and tracking
\cite{HMM} and to monitoring \cite{MON}). A characteristic
requirement of several wireless networks, which enables them to
adapt themselves to a changing environment, is that they be
``self-configuring," i.e., that a large number of wireless nodes
organize themselves to perform the tasks required by the
application they have been deployed for: examples of
self-configuration include construction of routing paths,
clustering, and formation of minimum-weight trees. In this paper,
we consider an aspect of self-configuration in wireless networks
referred to as {\em neighbor discovery} (ND). Neighbor discovery
is the determination of all nodes in the network a given node may
directly communicate with. Knowledge of neighbors is essential for
all routing protocols, medium-access control protocols, and
several other topology-control algorithms. Ideally, nodes should
discover their neighbors as quickly as possible, which will allow
nodes to save energy in their discovery phase. Also, rapid
discovery allows for other protocols (such as routing protocols)
to quickly start their execution. In addition, ND may also be the
solution for ``partner selection'' in cooperative wireless
networks. In fact, cooperation among users may carry advantages
only if the partners are chosen in a proper way: for example,
``decode-and-forward'' (DAF) protocols may suffer from cooperation
with weak users, thus failing in the goal of increasing the
diversity order~\cite{VWL}.

Recently, a number studies on ND algorithms have appeared (see,
e.g., \cite{ND1,EPH} and the references therein). Most of these
approach ND at a protocol level,  defining node \emph{A} to be a
neighbor of node \emph{B} if \emph{A} can exceed \emph{B}'s signal
to noise-ratio requirement: as a consequence, \emph{A} is inserted
in the neighbor list of \emph{B} based solely upon successful
reception, at node \emph{B}, of a packet sent by node \emph{A}.
Moreover, the Internet Engineering Task Force proposes to perform
Neighbor Discovery ``at IP Layer"~\cite{IP6}. The corresponding
protocol assumes a broadcast capability at physical layer, and a
MAC which handles contention. Now, ND algorithms for wireless
networks may not be contention-based when energy constraints are
tight: retransmission in the case of a collision costs energy,
which might be a resource at a premium. In this context, we
consider a transmission scheme which avoids collisions at
modulation level and is based on simultaneous transmission of
signatures. In principle, if the nodes' waveforms were orthogonal,
no collision would occur. In practice, these waveforms exhibit a
small correlation, which causes an interference. red In a
``standard'' network approach, signature transmission fails
whenever there is interference, because this causes a
``collision'' which may occur very often during a ND session if
there are many neighbors. Using multiuser detection, interference
does not cause collisions (these are in a sense ``automatically
resolved'') and can be controlled by multiuser-detection
algorithms.

ND can be performed in a supervised or unsupervised manner. In
supervised methods, there is a central controller (e.g., a leader
node) which processes the signal received from all nodes,
determines the network configuration, and communicates to all
nodes their neighbor lists. Supervised ND algorithms are expected
to cost a large amount of energy, and hence they should be
discarded for energy-limited networks, like sensor networks.
Unsupervised ND algorithms have no central controller, and each
node discovers its own neighbors. Another important issue in ND
problems is the timing aspect. In~\cite{zhang}, the
frame-synchronous assumption is justified by the presence in each
node of Global Positioning System (GPS) devices. In~\cite{EPH},
asynchronous algorithms are addressed, assuming that nodes can
synchronize at bit level (which is the assumption we make in the
following).

The goal of this work is to provide the foundations of signal
processing for ND in wireless networks. We consider an
unsupervised wireless network,  where ND is operated independently
of the regular exchange of packets in a frequency-flat Gaussian
multiple-access channel, shared by $K+1$ nodes which transmit,
synchronously and independently, a set of known signatures
according to the scheme advocated in~\cite{EPH}. Each node is
identified by its own unique signature, and every node keeps a
list of all the signatures of the network.\footnote{We do assume
that ND can occur in a separate channel of a mobile system. In a
sensor network, where energy is at a premium, it makes sense to
perform ND as soon as nodes are deployed. Other ND algorithms can
be based on higher layer protocols such as IP, but these might
seriously waste energy. Even if TCP/IP is in use (in a mobile
network, say), then allocating a small fraction of overall system
resources at the physical layer to ND might be better than running
ND periodically as an application program.} A node is called a
\emph{neighbor} of the reference node if its amplitude, received
by the latter, exceeds a preassigned \emph{activity threshold},
say $\tau_A$.\footnote{Note that this definition can be
generalized: for example, one may define a neighbor as one whose
power-to-interference plus noise ratio exceeds a given threshold.
In this paper we stick to a more restrictive definition, which
allows simpler algorithms.  In general, any information gleaned
through ND will help scheduling, or medium-access control, or
routing algorithms to be more efficient, because they will be
armed with neighbor information.} Moreover, nodes cannot transmit
and receive simultaneously on the same channel,\footnote{For
simplicity, we disregard the more general case of nodes that can
be in an {\em idle} state, i.e., they are neither receiving nor
transmitting.} and the maximum number of active nodes is fixed and
finite. We clarify that a neighbor relation between two nodes need
not be bidirectional, since each node discovers those nodes it can
receive from.
%The balance of this work is organized as follows. We describe
%signal model and the problem formulation in Section \ref{SEC2}. In
%Section \ref{SEC3} the optimum Bayesian algorithm for ND problem
%is described while Section \ref{SEC3c} mainly introduces the
%optimum non-Bayesian approach. Due to impractical complexity of
%the optimum algorithm sub-optimum solution are discussed in
%Section \ref{SEC4} with particular emphasis for Linear ND Tests
%(LNDT). In Section \ref{SEC5} a theoretical analysis for LNDT is
%given. Section \ref{SEC5a} introduces a fully blind methods based
%on Subspace approach using MUSIC algorithm. Section \ref{SEC6}
%contains some numerical results describing the algorithms
%developed above, while Section \ref{SEC7} concludes this paper.
%Appendix \ref{appendice1} derive the distribution of the
%transmission model while in Appendix \ref{appendice3} the
%performances of a simple Gaussian model are given.

The organization of this paper is the following. In
Section~\ref{SEC2} we provide a model for the physical aspects of
the networks, and we formulate our problem. ND algorithms are
introduced in Section~\ref{SEC3}, and analyzed in
Section~\ref{SEC4}. Section~\ref{SEC5} shows some numerical
results, while Section~\ref{SEC6} concludes the paper.

\section{Signal model and problem formulation}\label{SEC2}
Our scenario is based on the transmission scheme illustrated in
Fig.~\ref{ND1}, which corresponds to node $0$ searching its own
neighbors among four other nodes.~\footnote{We consider node $0$
to be the reference node. Since all nodes are at the same
hierarchical level, the same analysis applies to any node.} In
every time interval (``slot''), each node $i$, $i=0, 1, \ldots,
K$, transmits its own signature, independently of the other nodes,
with probability $\varepsilon_i$, while otherwise (and hence with
probability $1-\varepsilon_i$) it senses the channel. This
probability is actually designed as a part of the algorithm, and
it influences the ND algorithm performance, as we shall examine in
our analysis.
%%%%%%%%%%%%%%%%%%%%%%%%%%%%%%%%%%%%%%%%%%%%%%%%%%%%%%%%%%%%%%%%%%%%%%%%%%%

%%%%%%%%%%%%%%%%%%%%%%%%%%%%%%%%%%%%%%%%%%%%%%%%%%%%%%%%%%%%%%%%%%%%%%%%%%%

The ND algorithm runs in a finite period, called a \emph{discovery
session}, whose duration is denoted $T_D$. During $T_D$, every
active node transmits a number of signals containing one or more
copies of its signature. Each signal has duration $T=T_D/N$, with
$N$ the number of slots in the discovery session.  The network is
assumed to be unsupervised, which implies that all nodes are
independent and at the same hierarchical level: as a consequence,
the ND algorithm is run in parallel by all nodes. Under the
assumptions made in Section~\ref{SEC1}, the baseband
representation of the signal received by node $0$ in the time
interval $\big[ (n-1)T,nT \big)$, $n\in \{1, 2, \ldots, N\}$, is
\begin{equation}\label{observations-1}
y(t)= \left\{
\begin{array}{ll}
\sum_{k=1}^K \psi_{k,n} \alpha_k s_k\big(t-(n-1)T\big)+z(t) & \textrm{if $\psi_{0,n}=0$}\\
0 & \textrm{if $\psi_{0,n}=1$} \\
\end{array} \right.
\end{equation}
where $\alpha_k$ denotes the channel gain, i.e., the complex
amplitude of the signal received from node $k$ and assumed to be
constant during all the discovery session, $s_k(\cdot)$ is the
$k$th node signature, $\psi_{k,n}$ is a random variable taking
value $1$ if node $k$ is transmitting at time $n$, and value $0$
otherwise (so that ${\mathbb P}(\psi_{k,n}=1)=\varepsilon_k$), and
$z(t)$ is additive white complex Gaussian noise having spectral
density $2N_0$. We assume $\alpha_k$ to be modeled by a complex
circularly symmetric Gaussian random variable with variance
$2\sigma^2_k$. The signatures can be expressed as
\begin{equation}
s_k(t)=\sum_{l=1}^{L} s_{l,k} \phi \big(t-(l-1) T_c\big)/\sqrt{L}
\end{equation}
where $s_{l,k}\in \{-1,+1\}$ is the $l$th chip of the $k$th
signature, $L$ is the processing gain, $T_c=T/L$ is the chip
duration, and $\phi(\cdot)$ is the (unit-energy) chip
waveform.\footnote{The signatures are assumed to have unit
energy.} The slots devoted to channel sensing need not be
adjacent: however, due to our flat-fading assumption, we may
assume, without any loss of generality, a sensing phase of
\begin{equation}
M_0=\sum_{n=1}^N (1-\psi_{0,n})=N-\nu_0
\end{equation}
consecutive slots with intermittent other-users activity, with
$\nu_0$ the number of slots where node "0" is transmitting. Notice
that $M_0$ is random ($N$ is assumed fixed and node 0 has its own
activity factor $\varepsilon_0$), but the value it takes is known
to node $0$. Hence, in all subsequent derivations we refer to a
given value of $M_0$. Of course, we may adopt the silent phases of
node $0$ as a time scale, recasting~(\ref{observations-1}), with a
slight notational abuse,  in the form:\footnote{Notice that the
index $n$ refers to consecutive time slots, while $p$ refers to
the time scale defined by the silent phase of node "$0$".}
\begin{equation}\label{observations}
y(t)= \sum_{k=1}^K \psi_{k,p} \alpha_k
s_k\big(t-(p-1)T\big)+z(t)\;, \quad 0 \leq t \leq M_0T, \quad
p=1,2,\ldots,M_0
\end{equation} Our problem is now reduced
to determining the indexes $k$ such that $\{|\alpha_k|\}_{k=1}^K$
exceed an ``activity threshold'' $\tau_{\rm A}$, based on model
(\ref{observations}).

Since $z(t)$ is white Gaussian noise, the components of $y(t)$
orthogonal to the subspace spanned by the signatures are
irrelevant to our detection problem~\cite{VanTrees}. As a
consequence, we might in principle adopt the signatures
themselves, and their delayed versions, as an expansion basis for
such a subspace. Alternatively, we may use the $L-$dimensional
orthonormal basis \beq \bigcup_{\ell=0}^{L-1}\left \{ \phi
\big(t-\ell T_c-(p-1)T \big)\right \} \eeq to expand the signal in
the interval $\big[(p-1)T,pT\big)$. The two approaches are
obviously equivalent, but the latter is mandatory in situations
where the discovering node has no prior information as to the
signatures of other users: although we do not deal blind ND in
this paper, we choose this one due to its inherent flexibility.

Defining the scalar products
 \beq
 {\bf y}_{i,p}\triangleq \int_{(p-1)T}^{pT} y(t)\phi^*
\big(t-(i-1) T_c-(p-1)T\big) \; dt
 \eeq
 with $^*$ denoting
conjugation, we obtain a vector representation
$\mathbf{y}_p\triangleq [y_{1,p},y_{2,p}, \ldots, y_{L,p}]^T$
\footnote{The symbol $^T$ denotes transposition operation} of the
signal received in $\big[(p-1)T,pT\big)$:
\begin{equation}\label{mod}
\mathbf{y}_p=\sum_{k=1}^K \psi_{k,p} \alpha_k
\mathbf{s}_k+\mathbf{z}_p=\mathbf{S}\mathbf{\Psi}_p\boldsymbol{\alpha}+\mathbf{z}_p
\end{equation}
where $\mathbf{s}_k\triangleq \frac{1}{\sqrt{L}}[s_{1,k}, s_{2,k},
\ldots, s_{L,k}]^T$, $\mathbf{S}\triangleq [\mathbf{s}_1,
\mathbf{s}_2, \ldots, \mathbf{s}_K]$, $\mathbf{\Psi}_p\triangleq
\text{diag}(\psi_{1,p}, \psi_{2,p}, \ldots, \psi_{K,p})$,
$\boldsymbol{\alpha}\triangleq [\alpha_1, \alpha_2, \ldots,
\alpha_K]^T$, $\mathbf{z}_p\triangleq [z_{1,p}, z_{2,p}, \ldots,
z_{L,p}]^T$, and
\begin{equation}
z_{i,p}\triangleq  \int_{(p-1)T}^{pT} z(t)\phi^* \big(t-(i-1)
T_c-(p-1)T\big) \; dt
\end{equation}
The ND problem now consists of assessing, after observing the set
of $M_0$ vectors $\mathbf{y}_{1:M_{0}}\triangleq
\{\mathbf{y}_1,\ldots,\mathbf{y}_{M_{0}}\}$,
 which ones, among $|\alpha_1|$, \dots, $|\alpha_K|$, exceed the ``activity
threshold'' $\tau_A$.

\section{ND algorithms}\label{SEC3}

A sensible criterion for the selection of a ND algorithm consists
of minimizing the probability of choosing, among the $K$ network
nodes under scrutiny, an erroneous set of neighbors of node $0$.
Since there are $2^K$ such sets, each corresponding to one
hypothesis $H$, this error probability is minimized by the maximum
a posteriori (MAP) decision rule:
\begin{equation}\label{MAP}
\widehat{H}=\arg \max_{H} P(H)p( {\bf y}_{1:M_{0}} \mid H)
\end{equation}
where $P(H)$ is the a priori probability of hypothesis $H$, and
$p( {\bf y}_{1:M_{0}} | H)$ is the probability density of the
observations given $H$. As an example, if $K=2$, the $4$
hypotheses are shown in Table~\ref{tableH}.
%%%%%%%%%%%%%%%%%%%%%%%%%%%%%%%%%%%%%%%%%%%%%%%%
 \begin{table}[htbp]  \centering
 \begin{tabular*}{0.6\hsize}{@{\extracolsep{\fill}}cccc}\sphline
  $H_1$ & $H_2$ & $H_3$ & $H_4$ \\
  \hline\hline
  & & & \\
  $|\alpha_1|<\tau_A$ & $|\alpha_1|>\tau_A$ & $|\alpha_1|<\tau_A$ &
  $|\alpha_1|>\tau_A$\\
  & & & \\
  $|\alpha_2|<\tau_A$ & $|\alpha_2|<\tau_A$ & $|\alpha_2|>\tau_A$ & $|\alpha_2|>\tau_A$  \\
  & & & \\
  \hline
\end{tabular*}
\caption{\sl Hypotheses on the set of neighbors of node $0$ in a
network with $3$ nodes.} \label{tableH}
\end{table}
%%%%%%%%%%%%%%%%%%%%%%%%%%%%%%%%%%%%%%%%%%%%%%%%%%%%

Now, $p(\mathbf{y}_{1:M_{0}} \mid H)$ depends on the actual
pattern of transmit\slash receive intervals of each node, denoted
$\mathbf{\Psi}_{1:M_{0}}$. Since this is unknown under our
assumption that the transmission of signatures is not coordinated,
it should be obtained from the marginalization
\[
\sum_{\mathbf{\Psi}_{1:M_{0}}}P(\mathbf{\Psi}_{1:M_{0}})
p(\mathbf{y}_{1:M_{0}} \mid H, \mathbf{\Psi}_{1:M_{0}})
\]
which has a complexity that grows exponentially with $KM_0$.

To overcome this complexity obstacle, the decision on the neighbor
set works as follows. We first obtain estimates of the
instantaneous powers $\widehat{\left|{\alpha}_i\right|^2}$ of all
nodes, next we decide that a node is a neighbor by comparing each
of them with a threshold, i.e.,
 \beq\label{test}\widehat{
\left|\alpha_i\right|^2} \threshold \tau_i^2
 \eeq
 where
\begin{description}
\item[$H_1$:] The received instantaneous power exceeds $\tau^2_A$.
\item[$H_0$:] The received instantaneous power is below
$\tau^2_A$.
\end{description}

The performance of this test can be expressed through its
probability $P_F^{(i)}$ of a {\em false-alarm} and its probability
$P_M^{(i)}$ of a {\em miss}, defined as:
 \beq\label{PF-PM}
\begin{array}{lll}
P_F^{(i)}& = & {\mathbb P} \left\{
\widehat{|\alpha_i|^2}>\tau^2_i\mid |\alpha_i|<\tau_A \right\}
\\
P_M^{(i)}& = & {\mathbb P} \left\{
\widehat{|\alpha_i|^2}<\tau^2_i\mid |\alpha_i|>\tau_A \right\}
\end{array}
\eeq These are related to the overall error probability
through\footnote{In what follows, the superscripts will be skipped
whenever no confusion is induced by this notational
simplification.}
  \beq\label{Pe} P^{(i)}(e)=P_F^{(i)} {\mathbb P}
\left\{|\alpha_i|<\tau_A\right\}+P_M^{(i)} {\mathbb P}
\left\{|\alpha_i|>\tau_A\right\}
  \eeq

Now, the maximum-likelihood (ML) estimators of the instantaneous
powers can be obtained by jointly estimating $\boldsymbol{\alpha}
$ and the matrix sequence $\mathbf{\Psi}_{1:M_{0}}$.
Straightforward calculations show that the ML estimates of
$\boldsymbol{\alpha} $ and $\bPsi_{1:M_{0}}$ result from the
solution of the $2^{KM_{0}}$ linear systems --- each corresponding
to an outcome $\bPsi_{1:M_{0},i}$ of the matrix sequence
$\bPsi_{1:M_{0}}$:
\begin{equation} \left(\sum_{p=1}^{M_{0}}
\bPsi_{p,i}\bS^\dagger \bS\bPsi_{p,i}
\right)\balpha=\left(\sum_{p=1}^{M_{0}} \bPsi_{p,i}\bS^\dagger
\by_p \right)\;. \label{eq:linear}
\end{equation}
with $^\dagger$ denoting Hermitian operation. Computing \beq
\widehat{\balpha}_{ML}=\arg \min_{i=1,\ldots,2^{KM_{0}}}
\sum_{p=1}^{M_{0}}\parallel
\by_p-\bS\bPsi_{p,i}\widehat{\balpha}_i
\parallel^2\eeq
with $\widehat{\balpha}_i$ the solution corresponding to
$\bPsi_{1:M_{0},i}$, and recalling that ML estimates commute under
nonlinear transformations, test (\ref{test}) can be implemented by
using $\widehat{
\left|\alpha_i\right|^2}=|\widehat{\alpha}_{ML,i}|^2$,

Even with this receiver, implementation complexity would be
unrealistic, and hence a further simplification is called for.
Instead of dealing with the receive\slash transmit pattern related
to the whole discovery session, we rather obtain estimates based
on a single $T$-interval observation, which are then combined
according to a suitable integration strategy.

\subsection{Suboptimum ND algorithms}

Consider again model~(\ref{mod}). The ML estimate of
$\bPsi_p\balpha$, based upon the observation $\by_p$ available in
slot $p$, is \beq\label{MLE} \widehat{\bPsi_p
\balpha}=(\bS^\dagger \bS)^{-1} \bS^\dagger  \by_p=\bS^+ \by_p
\eeq where $\bS^+$ denotes the pseudo-inverse of the tall matrix
$\bS$.

A closer look at this solution reveals that, since
 \beq
\bS^+\by_p=\bS^+\bS\bPsi_p\balpha+\bS^+\bz_p=\bPsi_p\balpha+\bw_p
\label{eq:filtered} \eeq
 with $\E [\bw_p\bw_p^\dagger ]=2\N(\bS^\dagger \bS)^{-1}$, the interference from the
other users is completely eliminated, at the price of some noise
enhancement, reflecting the increase of the variance of its $i$th
component by the factor $\{(\bS^\dagger
\bS)^{-1}_{i,i}\}_{i=1}^{K}$. It is interesting to notice that
this estimate is noise-limited, but not interference-limited,
implying that any receiver based on (\ref{MLE}) is {\em
asymptotically efficient} \cite{VRD}; likewise, {\em near-far
resistance} is granted \cite{VRD}.

Since there are $M_0$ sensing phases,  the $M_0$ estimates
resulting from repeated application of (\ref{MLE}) should be
combined to yield the final test statistic. Borrowing techniques
from radar detection theory, reasonable combination criteria are
{\em coherent integration} (CI), wherein an estimate of the
instantaneous power is obtained as
 \beq\label{coherent}
\widehat{|\alpha_{i}|^2}_{CI}\triangleq
\left|\frac{1}{M_0}\sum_{p=1}^{M_0}(\bS^{+}\by_p)_i\right|^2 \eeq
and {\em incoherent integration} (II)
 \beq\label{incoherent}
\widehat{|\alpha_{i}|^2}_{II}\triangleq
\frac{1}{M_0}\sum_{p=1}^{M_0}|(\bS^{+}\by_p)_i|^2 \eeq
 Notice that
\begin{eqnarray}
\E \left[\left.\widehat{
|\alpha_{i}|^2}_{CI}\right|M_0,|\alpha_i|^2\right]&=&
%\E[\nu_i^2]|\alpha_i|^2/M+2\N(\bS^\dagger \bS)^{-1}_{i,i}/M=
\varepsilon_i|\alpha_i|^2\left[\varepsilon_i+\frac{1-\varepsilon_i}{M_0}\right]
+\frac{2\N(\bS^\dagger \bS)^{-1}_{i,i}}{M_0}\label{estimator_CI}\\
\E\left[\left.\widehat{
|\alpha_{i}|^2}_{II}\right|M_0,|\alpha_i|^2\right]&=&
%\E[\nu_i]|\alpha_i|^2/M+2\N(\bS^\dagger \bS)^{-1}_{i,i}=
|\alpha_i|^2\varepsilon_i+2\N(\bS^\dagger
\bS)^{-1}_{i,i}\label{estimator_II}
%\mbox{Var}[\widehat{|\alpha_i|^2}|M]=\frac{\mbox{Var}[|\alpha_i|^2]\varepsilon+
%4\N^2\left((\bS^\dagger \bS)^{-1}_{i,i}\right)^2+4\varepsilon\E[|\alpha_i|^2]2\N(\bS^\dagger \bS)^{-1}_{i,i}}{M}
%(1-\varepsilon)}{M}+\frac{2\N N
%\varepsilon}{M}
\end{eqnarray}
implying that both $\widehat{|\alpha_{i}|^2}_{II}$ and
$\widehat{|\alpha_{i}|^2}_{CI}$ can be interpreted as biased
estimators of the instantaneous power received in each slot from
node $i$: biases can however be absorbed in the detection
thresholds $\tau_i$, while what matters here is that they are both
{\em consistent} in the mean square sense, a property that will be
exploited later on. Inserting (\ref{coherent}) and
(\ref{incoherent}) into (\ref{test}), and skipping factors that
can be absorbed in the detection thresholds, we obtain the {\em
coherent detector} (CD)
 \beq \left\{
\begin{array}{l} \left|\sum_{p=1}^{M_0}(\bS^{+}\by_p)_i\right|^2 >
\tau_i^2 \rightarrow \mbox{node $i$ is a neighbor}
\\ \left|\sum_{p=1}^{M_0}(\bS^{+}\by_p)_i\right|^2 < \tau_i^2
\rightarrow \mbox{node $i$ is not a neighbor} \end{array} \right.
\label{CD} \eeq and the Incoherent Detector (ID):\beq \left\{
\begin{array}{l} \sum_{p=1}^{M_0}|(\bS^{+}\by_p)_i|^2 >
\tau_i^2 \rightarrow \mbox{node $i$ is a neighbor}
\\\sum_{p=1}^{M_0} |(\bS^{+}\by_p)_i|^2 < \tau_i^2
\rightarrow \mbox{node $i$ is not a neighbor} \end{array} \right.
\label{TSML} \eeq

Notice how the CD can also be interpreted in a different way.
Indeed, it may be obtained by first pre-processing the
observations so as to form the cumulative sum:  \beq \mathbf{y}
\triangleq \sum_{p=1}^{M_0} \mathbf{y}_p=\sum_{p=1}^{M_0} \bigg(
\sum_{k=1}^K \psi_{k,p} \alpha_k \mathbf{s}_k+\mathbf{z}_{p}\bigg)
=\sum_{k=1}^K \nu_k \alpha_k
\mathbf{s}_k+\mathbf{z}=\mathbf{S}\mathbf{V}\boldsymbol{\alpha}+\mathbf{z}\label{sig_mod_sum}
\eeq
 where
  \beq \nu_k \triangleq\sum_{p=1}^{M_0}\psi_{k,p} \qquad
\mathbf{z}\triangleq \sum_{p=1}^{M_0} \mathbf{z}_{p}
 \eeq
  and $\mathbf{V}\triangleq \text{diag}(\nu_1, \ldots, \nu_K)$, then multiplying
the new observation by $\bS^{+}$ and finally extracting the $i$th
component to form the test statistic (\ref{CD}). Rewriting
equation (\ref{sig_mod_sum}) in the form:
\begin{equation}
\mathbf{y}=\underbrace{\nu_i\alpha_i\mathbf{s}_i}_{\text{useful
signal}}+\underbrace{\sum_{k\neq i}\nu_k\alpha_k
\mathbf{s}_k}_{\text{interference}}+\underbrace{\mathbf{z}}_{\text{noise}}
\end{equation}
with $\mathbf{z}\sim \mathcal{N}_c(0,2\N M_0 \mathbf{I}_L)$, where
$\mathbf{I}_L$ is the $L \times L$ identity matrix, the CD is
easily seen to be a member of the family of {\em linear ND tests}
(LNDT), wherein a decision on the proximity of user  $i$ is made
based on the rule:
\begin{equation}\label{LINT}
|\mathbf{c}_i^\dagger \mathbf{y}|^2 \threshold \tau_i^2
\end{equation}
Thus, the CD (\ref{CD}) can be also interpreted as the
zero-forcing (ZF) member of the family (\ref{LINT}), obtained as
the unique solution to the constrained minimization problem:
\begin{equation}
\left\{
\begin{array}{ll}
\mathbf{c}_{i,ZF}=\arg \min _{\mathbf{c}_i}\text{E}\big[
|\mathbf{c}_i^\dagger  \sum_{k=1}^K \alpha_k \nu_k \mathbf{s}_k|^2
\big] &
\\ \mathbf{c}_{i,ZF}^\dagger \mathbf{s}_i=\beta^2 & \\
\end{array} \right.
\end{equation}
with $\beta \neq 0$, which yields\footnote{Notice from (\ref{ZF})
that the parameter $\beta^2$ has been set to
$\frac{1}{(\bS^\dagger \bS)^{-1}_{i,i}}$}
 \beq\label{ZF}
\mathbf{c}_{i,ZF}=\left(\mathbf{I}_L-\mathbf{S}_i\mathbf{S}_i^{+}\right)\mathbf{s}_i={\cal
P }_i \mathbf{s}_i
 \eeq
 where $\mathbf{S}_i$ is the $L \times (K-1)$ matrix
obtained skipping the $i-$th column from $\mathbf{S}$ and ${\cal P
}_i$ denotes the projector onto the orthogonal complement of the
column span of $\mathbf{S}_i$. For future reference we remind here
that \cite{VRD}\beq\label{orthogonal} |\mathbf{c}_{i,ZF}^\dagger
\mathbf{s}_k|^2=\left\{
\begin{array}{ll}0 & \mbox{if $k \neq i$}\\
|\mathbf{s}_i^\dagger {\cal P }_i\mathbf{s}_i|^2=\parallel
\mathbf{s}_{i,\perp}\parallel^4& \mbox{if $k=i$}
\end{array}
\right.
 \eeq
 where $\mathbf{s}_{i,\perp}$ denotes the projection of
 $\mathbf{s}_{i}$ on the above orthogonal complement: needless to say, since
 $\parallel \mathbf{s}_{i,\perp}\parallel^2=1/[(\bS^\dagger \bS)^{-1}_{i,i}]$,  the noise power is enhanced by a factor
 $(\bS^\dagger \bS)^{-1}_{i,i}$.

The vector $\mathbf{c}_i$ can be designed according to a number of
different criteria. For example, in \cite{ABL} an LNDT based on
conventional matched filtering (MF), i.e., assuming
\begin{equation}\label{conv_filt}
\mathbf{c}_{i,MF}\triangleq \mathbf{s}_i
\end{equation}
has been proposed and analyzed for ND~\cite{ABL}. MF is indeed
simple, but it results into interference-limited performance, as
we shall prove soon, nor does it retain the near-far resistance
property granted by ML-based detectors.

A possible alternative to the ZF criterion is offered by the
minimum-mean-output-energy (MMOE) strategy, first introduced
in~\cite{HMV}, wherein the vector $\mathbf{c}_i$ is obtained as
the unique solution to the following constrained minimization
problem:
\begin{equation}
\left\{
\begin{array}{ll}
\mathbf{c}_{i,MMOE}=\arg \min _{\mathbf{c}_i}\text{E}\bigg[
\bigg|\mathbf{c}_i^\dagger  \bigg(\sum_{k=1}^K \alpha_k \nu_k
\mathbf{s}_k+\mathbf{n}\bigg)\bigg|^2 \bigg] & \\
\mathbf{c}_{i,MMOE}^\dagger \mathbf{s}_i=1 & \\
\end{array} \right.
\end{equation}
namely:
\begin{equation}
\mathbf{c}_{i,MMOE}=\frac{\mathbf{M}_{\mathbf{y}\mathbf{y}}^{-1}
\mathbf{s}_i}{\mathbf{s}_i^\dagger
\mathbf{M}_{\mathbf{y}\mathbf{y}}^{-1}\mathbf{s}_i}
\end{equation}
where $\mathbf{M}_{\mathbf{y}\mathbf{y}}\triangleq \sum_{k=1}^K
2\sigma_k^2 E[\nu_k^2] \mathbf{s}_k \mathbf{s}_k^\dagger +2\N
M_0\mathbf{I}_L$.  Due to the invariance of the decision rule to
any positive scaling of the test statistic, an equivalent detector
relies upon setting
\begin{equation}\label{MMOE}
\mathbf{c}_{i,MMOE}=\mathbf{M}_{\mathbf{y}\mathbf{y}}^{-1}
\mathbf{s}_i
\end{equation}
It might be worth recalling here that, since \beq \lim_{\N
\rightarrow 0} \mathbf{M}_{\mathbf{y}\mathbf{y}}^{-1} \mathbf{s}_i
\propto {\cal P }_i \mathbf{s}_i\eeq MMOE is itself asymptotically
efficient. Likewise, it retains the near-far resistance property
since the projection direction $\mathbf{c}_{i,MMOE}$ tends to
become orthogonal to those signatures whose amplitudes become
increasingly large \cite{HMV}. The advantage of (\ref{MMOE}) over
ZF is that it easily lends itself to adaptive implementations in
situations where the signatures of the active users are unknown.
Even though we do not deal with adaptive ND in this paper, we
anticipate that a number of reduced complexity algorithms, ranging
from the ${\cal O}(L)$-complex Least Mean Squares to the ${\cal
O}(L^2)$-complex Recursive Least Squares, can be easily applied
for adaptive MMOE implementation.

\section{Analysis}\label{SEC4}
From now on we assume that the node to be detected is node "1".
Consider first the ID. The conditional false-alarm and miss
probabilities in assessing the proximity of node $1$ can be
written as:
\begin{eqnarray}
P_M&=&\mathbb{P}(\chi_1<\tau_1^2\big|
|\alpha_1|>\tau_A,\mathbf{\Psi}_{1:M_{0}})\\
P_F&=&\mathbb{P}(\chi_1>\tau_1^2\big|
|\alpha_1|<\tau_A,\mathbf{\Psi}_{1:M_{0}})
\end{eqnarray}
with $\chi_1\triangleq\sum_{p=1}^{M_{0}} |(\mathbf{S}^+
\mathbf{y}_p)_1|^2$. Given $|\alpha_1|$ and
$\mathbf{\Psi}_{1:M_{0}}$, $\chi_1$ is noncentral chi-square
distributed with $2M_0$ degrees of freedom and parameters $\nu_1
|\alpha_1|^2$ and $\sigma^2_{n,1}=(\mathbf{S}^\dagger
\mathbf{S})^{-1}_{1,1}\N$, implying
\begin{equation}
\mathbb{P}(\chi_1>\tau_1^2 \mid  |\alpha_1|,
\mathbf{\Psi}_{1:M_{0}})=Q_{M_0}\bigg(\frac{\sqrt{\nu_1}|\alpha_1|}{\sigma_{n,1}},\frac{\tau_1}{\sigma_{n,1}}
\bigg)
\end{equation}
where $Q_{M_{0}}(\cdot, \cdot)$ is the Marcum function of order
$M_0$. Using the series expansion of modified Bessel functions
 \beq
I_n(x)= \sum_{k=0}^{\infty} \frac{(x/2)^{n+2k}}{k! \Gamma(n+k+1)}
 \eeq
 we obtain
\begin{equation}\label{joint_p} \begin{array}{c}
\mathbb{P}(\chi_1>\tau_1^2 \mid  |\alpha_1|,
\mathbf{\Psi}_{1:M_{0}})=e^{-\frac{|\alpha_1|^2\frac{\nu_1}{\sigma^2_{n,1}}}{2}}\sum_{k=0}^\infty
\frac{\left(|\alpha_1|\sqrt{\frac{\nu_1}{\sigma^2_{n,1}}}\right)^{2k}}{2^k
k! \Gamma(M_0+k)}\Gamma\bigg(M_0+k;
\frac{\tau^2_1}{2\sigma^2_{n,1}}\bigg)\\
\mathbb{P}(\chi_1>\tau_1^2, |\alpha_1|>\tau_A
\big|\mathbf{\Psi}_{1:M_{0}})=\frac{1}{1+\nu_1\rho_1}\sum_{k=0}^\infty
\bigg( \frac{\nu_1 \rho_1}{1+\nu_1 \rho_1}\bigg)^k
Q\bigg(M_0+k;\frac{\tau_1^2}{2\sigma^2_{n,1}}\bigg)Q\bigg(k+1;\frac{\tau_A^2}{2\sigma^2_1}(1+\nu_1\rho_1)\bigg)
\end{array}
\end{equation}
where
\begin{equation}
\Gamma(k;x)\triangleq \int_{x}^{\infty} t^{k-1} e^{-t} dt \; ,
\qquad Q(k;x)=\frac{\Gamma(k;x)}{\Gamma(k)}
\end{equation}
are the upper incomplete Gamma function and its regularized
version, respectively, while $\rho_1$ is the signal-to-noise ratio
after decorrelation, i.e.: \beq
\rho_1=\frac{\sigma^2_1}{\sigma^2_{n,1}}=\frac{\sigma^2_1}{\N(\bS^\dagger
\bS)^{-1}_{1,1}} \eeq We thus obtain the conditional measure:
%are the upper and lower incomplete Gamma functions, respectively.
\begin{equation}\label{PM-inco}
P_M=1-\frac{ e^{\frac{\tau_A^2}{2\sigma^2_1}}}   {1+\nu_1
\rho_1}\sum_{k=0}^\infty \bigg( \frac{\nu_1 \rho_1}{1+\nu_1
\rho_1}\bigg)^k
Q\bigg(M_0+k;\frac{\tau_1^2}{2\sigma^2_{n,1}}\bigg)Q\bigg(k+1;\frac{\tau_A^2}{2\sigma^2_1}(1+\nu_1\rho_1)\bigg)
\end{equation}
which should be averaged over $\nu_1$ to yield the conditional
probability of a miss given $M_0$. Similar developments hold for
$P_F$, yielding \beq\label{P_F_G}
\mathbb{P}(\chi_1>\tau_1^2,|\alpha_1|<\tau_A\big|
\mathbf{\Psi}_{1:M_{0}})=\frac{1}{1+\nu_1\rho_1}\sum_{k=0}^\infty
\bigg( \frac{\nu_1 \rho_1}{1+\nu_1 \rho_1}\bigg)^k
Q\bigg(M_0+k;\frac{\tau_1^2}{2\sigma^2_{n,1}}\bigg)
P\bigg(k+1;\frac{\tau_A^2}{2\sigma^2_1}(1+\nu_1\rho_1)\bigg) \eeq
where
\begin{equation}
 \gamma(k;x)\triangleq
\int_{0}^{x} t^{k-1} e^{-t} dt\; , \qquad
P(k;x)=\frac{\gamma(k;x)}{\Gamma(k)}
\end{equation}
are the lower incomplete Gamma function and its regularized
version, respectively. Finally, from (\ref{P_F_G}) we easily
obtain:
\begin{equation}\label{PF-inco}
P_F=\frac{ e^{\frac{\tau_A^2}{4\sigma^2_1}}}{2(1+\nu_1 \rho_1)}
\text{csch}\bigg(\frac{\tau_A^2}{4\sigma^2_1}
\bigg)\sum_{k=0}^\infty \bigg( \frac{\nu_1 \rho_1}{1+\nu_1
\rho_1}\bigg)^k
Q\bigg(M_0+k;\frac{\tau_1^2}{2\sigma^2_{n,1}}\bigg)
P\bigg(k+1;\frac{\tau_A^2}{2\sigma^2_1}(1+\nu_1\rho_1)\bigg)
\end{equation}

Consider now the test family (\ref{LINT}). Notice that, since
\begin{equation}
g_1\triangleq\mathbf{c}_1^\dagger
\mathbf{y}=\underbrace{\nu_1\mathbf{c}_1^\dagger \mathbf{s_1}
\alpha_1}_{\text{useful
signal}}+\underbrace{\sum_{k=2}^K\nu_k\mathbf{c}_1^\dagger
\mathbf{s_k} \alpha_k+\mathbf{c}_1^\dagger
\mathbf{z}}_{\text{interference+noise}}
\end{equation}
$|g_1|^2$ is conditionally chi-square with two degrees of freedom,
given $\alpha_1$, $\{\nu_i\}_{i=1}^{K}$ and $M_0$, with
non-centrality parameter $|\nu_1\mathbf{c}_1^\dagger \mathbf{s_1}
\alpha_1|^2=|\alpha_1|^2\nu_1^2\mathbf{c}_1^\dagger
\mathbf{s_1}\mathbf{s_1}^\dagger \mathbf{c}_1$ and scale parameter
\begin{equation}
\Sigma^2(\mathbf{c}_1) \triangleq \sum_{k=2}^K
|\nu_k\mathbf{c}_1^\dagger
\mathbf{s_k}|^2\sigma_k^2+M_0\N\|\mathbf{c}_1\|^2= \sum_{k=2}^K
\sigma_k^2\nu_k^2\mathbf{c}_1^\dagger
\mathbf{s_k}\mathbf{s_k}^\dagger
\mathbf{c}_1+M_0\N\|\mathbf{c}_1\|^2
\end{equation}
whereby, reproducing the same steps leading to (\ref{PM-inco}) and
(\ref{PF-inco}), we obtain:
\begin{eqnarray}
P_M&=&1-\frac{ e^{\frac{\tau_A^2}{2\sigma^2_1}}}   {1+\nu_{1}^2
\rho_{eq}}\sum_{k=0}^\infty \bigg( \frac{\nu_1^2
\rho_{eq}}{1+\nu_1^2 \rho_{eq}}\bigg)^k
Q\bigg(k+1;\frac{\tau_1^2}{2\Sigma^2(\mathbf{c}_1)}\bigg)Q\bigg(k+1;\frac{\tau_A^2}{2\sigma^2_1}(1+\nu_1^2\rho_{eq})\bigg)\label{PM-lin}\\
P_F&=&\frac{ e^{\frac{\tau_A^2}{4\sigma^2_1}}}{2(1+\nu_1^2
\rho_{eq})} \text{csch}\bigg(\frac{\tau_A^2}{4\sigma^2_1}
\bigg)\sum_{k=0}^\infty \bigg( \frac{\nu_1^2 \rho_{eq}}{1+\nu_1^2
\rho_{eq}}\bigg)^k
Q\bigg(k+1;\frac{\tau_1^2}{2\Sigma^2(\mathbf{c}_1)}\bigg)
P\bigg(k+1;\frac{\tau_A^2}{2\sigma^2_1}(1+\nu_1^2\rho_{eq})\bigg)\nonumber\\
&&\label{PF-lin}
\end{eqnarray}
where $\rho_{eq}$ represents the signal-to-interference-plus-noise
ratio (SINR) at the output of the linear filter, i.e.: \beq
\rho_{eq}=\frac{\sigma^2_1\mathbf{c}_1^\dagger
\mathbf{s_1}\mathbf{s_1}^\dagger
\mathbf{c}_1}{\Sigma^2(\mathbf{c}_1)}\eeq Relationships
(\ref{PM-lin}) and (\ref{PF-lin}) are quite reminiscent of
(\ref{PM-inco}) and (\ref{PF-inco}), respectively, one major
difference being the dependency of the performance on
$\nu^2_1\rho_{eq}$, rather than $\nu_1\rho_{1}$. Of course, the
quadratic factor in $\nu_1$ stems from the fact that linear
detectors operate on a coherent combination of the observations,
while ID combines the slot-by-slot estimates incoherently. Notice,
however, that the above relationships represent conditional
measures, given $M_0$ (i.e., given $\nu_0$) and
$\{\nu_i\}_{i=1}^K$.
%On the other hand, ID performs, at each step, a complete
%elimination of the MAI, through the pseudo-inversion in
%(\ref{MLE}), whereby its performance depends on the
%signal-to-noise ratio only.
If the discovery session is long enough, so that the matrix
sequence $\bPsi_{1:M_{0}}$ may exhibit its typical behavior,
namely, if $N(1-\varepsilon_0)\gg 1$,  then the $\nu_k$'s tend in
probability to $M_0\varepsilon_k$, whereby the unconditional
performances may be obtained by averaging the corresponding
conditional measures on the typical set of values of
$\{\nu_k\}_{k=1}^{K}$ and $M_0$ only, implying:
\begin{description}
\item{-} $M_0 \simeq N(1-\varepsilon_0)$; \item{-} $\nu_k \simeq
M_0\varepsilon_k=N\varepsilon_k(1-\varepsilon_0)$.
\end{description}
In this limiting situation, it is interesting to notice the
relationship between the "cumulated" SNR's for ID and CD (i.e.,
the ZF of (\ref{ZF})), i.e. (see also (\ref{orthogonal}) and
subsequent comments): \beq \label{SNRDEC}
\nu_1^2\rho_{eq}=\frac{\sigma^2_1\nu_1^2\parallel
\mathbf{s}_{1,\perp}\parallel^4}{M_0\N\parallel
\mathbf{s}_{1,\perp}\parallel^2}=\frac{\nu_1^2\sigma^2_1}{M_0\N(\bS^\dagger
\bS)^{-1}_{1,1}} \simeq \varepsilon_1\nu_1 \rho_1 \eeq Thus, in
terms of cumulated signal-to-noise ratio and for large $N$, ID
seems to be preferable to CD, even though a global superiority
cannot be claimed due to the different forms assumed by the
respective false-alarm and miss probabilities.

So far no criterion has been given to select the decision
threshold $\tau_1$. Notice, however, that the consistency of the
estimates (\ref{coherent}) and (\ref{incoherent}) allows devising
the asymptotically optimum  thresholds (those achieving minimum
error probability for large $N$) from (\ref{estimator_CI}) and
(\ref{estimator_II}) in the form:
\begin{eqnarray} \tau^2_{1,CD}&=& N(1-\varepsilon_0)\left[
\varepsilon_1
\tau_A^2\left[N(1-\varepsilon_0)\varepsilon_1+(1-\varepsilon_1)\right]+2\N(\bS^\dagger
\bS)^{-1}_{1,1}\right]
  \label{tauCD}\\
\tau^2_{1,ID}&=&N(1-\varepsilon_0) \left[
\tau_A^2\varepsilon_1+2\N(\bS^\dagger \bS)^{-1}_{1,1}\right]
\label{tauID}
\end{eqnarray}
For short discovery sessions, and under known activity factors of
nodes to be discovered, optimum detection thresholds can be
obtained by  evaluating numerically the unconditional error
probability, and then determining the points where it has a
minimum.

\section{Results}\label{SEC5}
We consider here a fully loaded network with $K+1=7$, each node
being assigned a length-$7$ $m$-sequence. As in the previous
section, we assume that node "0" has to decide on the proximity of
node $1$. Figure \ref{FIG1} assumes SNR$_1\triangleq
\sigma_1^2/\N=0$~dB, $N=100$, a power-controlled scenario wherein
all nodes are received with the same average power, uniform
activity factor ($\varepsilon_k=\varepsilon=0.5$), and an activity
threshold equal to the median of the fading amplitude
distribution, i.e., such that $\mathbb{P}(|\alpha_1|>\tau_A)=0.5$.
The figure represents the pair $P_M$, $P_F$ for the various
receivers examined so far. Interestingly, ``conventional'' MF
suffers from the presence of the other nodes even in this rather
benign situation, while MMOE, ZF (which coincides with CD rule)
take advantage of their asymptotic efficiency: from now on,
conventional MF will not be considered any longer. The interested
reader is deferred to \cite{ABL} for a fairly thorough performance
assessment.

The reliability of the asymptotic approximation for long discovery
sessions can be assessed through figures \ref{FIG2a}--\ref{FIG2c}
for the CD, and through figures \ref{FIG3a}--\ref{FIG3c} for the
ID, which refer to the same scenario as in Fig.~\ref{FIG1}. The
curves of these figures represent:
\begin{description}
\item{-} The unconditional false alarm and miss probabilities
obtained by simulation.
\item{-} The same pair obtained by a
semi-analytical method, i.e., by estimating the averages of their
conditional counterparts.
\item{-} The asymptotic approximation.
\end{description}
From the plots, it is evident that the asymptotic approximation
tends to overestimate the performances in the interesting region
of low error probabilities,
%- a consequence of the
%Jensen's inequality applied to the functions $P_M(\nu_1)$ and
%$P_F(\nu_1)$, locally concave in the region $P_M(\nu_1)
%\rightarrow 0$ $P_F(\nu_1) \rightarrow 0$ -
while coming closer and closer to the true performance as $N$
increases: notice that the approximation is extremely tight for
$N=500$, a realistic value indeed in real applications, which, for
$\epsilon_0=0.5$, corresponds to $M_0\simeq250$. However, it
should be kept in mind that, for larger activity factors of the
discovering node, the minimum value of $N$ for the asymptotic
behavior to be reached inevitably increases.

The validity of the approximations (\ref{tauCD}) and
(\ref{tauID}), yielding the asymptotically optimal thresholds for
CD and ID, respectively, can be verified through Figs.~\ref{FIG4}
and \ref{FIG5}, showing $P(e)$ versus $\tau_1$ for CD and ID,
respectively, for some values of SNR$_1$, $\varepsilon$=0.5,
$N=500$ and $\tau_A$ such that
$\mathbb{P}(|\alpha_1|>\tau_A)=0.5$.
%\begin{eqnarray}
%\tau_{1,opt,DEC}&=&\sqrt{N(1-\epsilon)(N\epsilon^2(1-\epsilon)\tau_A^2+2N_0(\mathbf{S}^\dagger
%\mathbf{S})_{1,1})}\nonumber\\
%\tau_{1,opt,TSML}&=&\sqrt{N(1-\epsilon)(\epsilon\tau_A^2+2N_0(\mathbf{S}^\dagger
%\mathbf{S})_{1,1})}\nonumber
%\end{eqnarray}
These plots, obtained numerically, show that the error probability
admits a unique minimum; moreover, the optimal values of $\tau_1$
are surprisingly close to those resulting from the asymptotic
approximations in (\ref{tauCD}) and (\ref{tauID}). It might be
marginally worth noticing that such an optimum value is
practically independent of
SNR$_1$ for CD, while being strongly tied to SNR$_1$ for ID.

Fig.~\ref{FIG6} is aimed at comparing CD and ID. It represents the
error probability versus the signal-to-noise ratio SNR$_1$ using
the optimal thresholds for both receivers, and assuming again
$\varepsilon=0.5$, $N=500$, and $\tau_A$ as before. It is
interesting to notice that CD outperforms ID for small
signal-to-noise ratios, while ID is preferable for medium-to-large
values of SNR$_1$.
%%%%%%%%%%%%%%%%%%%%%%%%%%%%%%%%%%%%%%%%%%%%%%%%%%%%%%%%%%%%%%%%%%%

\section{Conclusions}\label{SEC6}
We have examined the problem of discovery which nodes are
neighbors in a wireless network operating over a fading channel.
The optimum Bayesian decision rule has been derived, showing that
its complexity is practically prohibitive. Two suboptimum
neighbor-discovery algorithms have been introduced, based on
standard techniques of coherent and incoherent integration. We
show how coherent integration may be viewed as a particular case
of a family of algorithm akin to Linear Neighbor Discovery Tests
(LNDT). Theoretical analysis allows one to understand the design
of a system employing such algorithms according to constraints on
error rate, signal-to-noise ratio and discovery session duration.
Finally, algorithm optimization was considered, and formulas were
derived for asymptotical optimum threshold.

\section{Acknowledgments}
The authors wish to express their gratitude to Anthony Ephremides
and Steven Borbash for useful discussions about their ND
algorithm~\cite{EPH}.

\newpage

 \insertfig{t}{0.7}{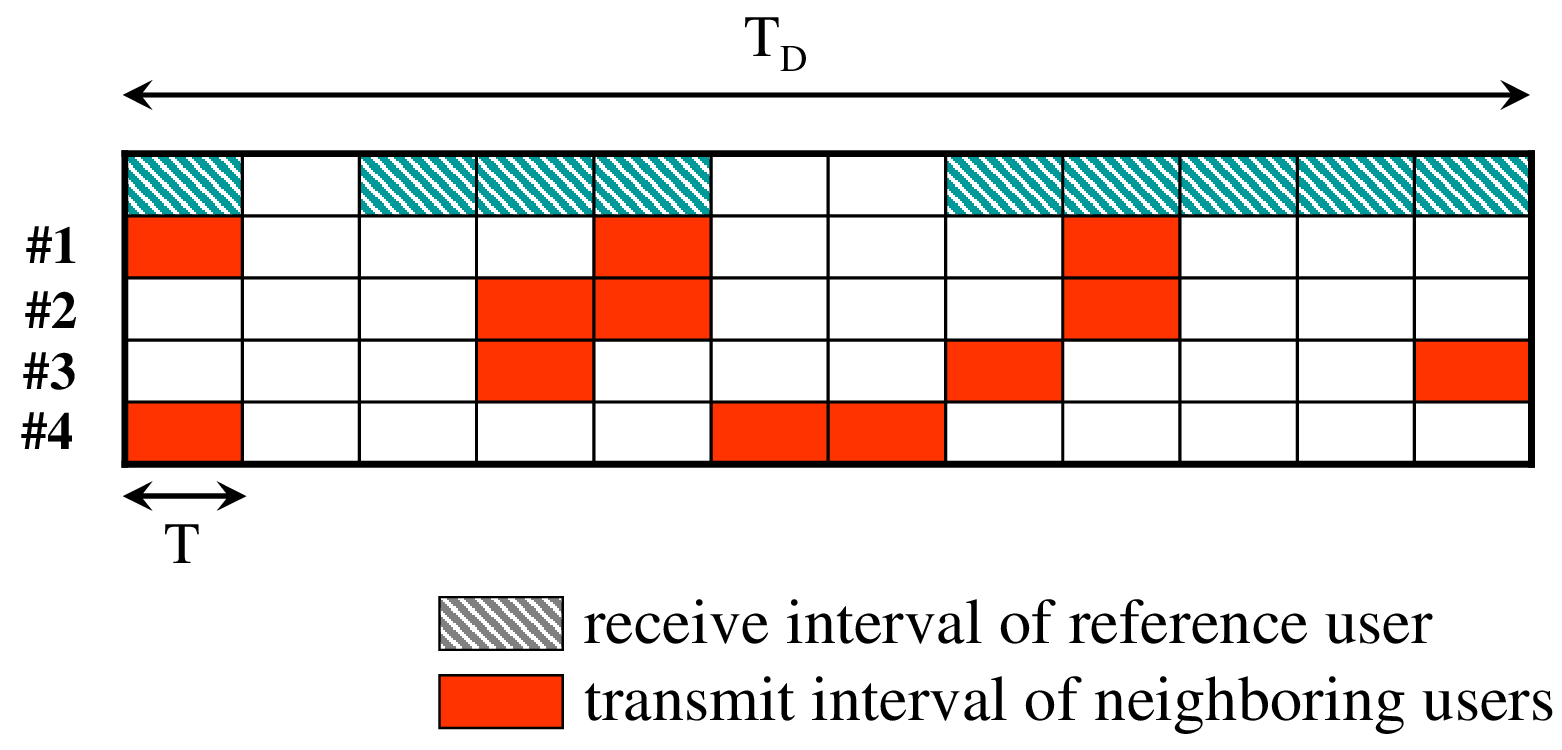}{\sl A scheme for synchronous neighbor
discovery.}{ND1}
 \insertfig{h}{0.8}{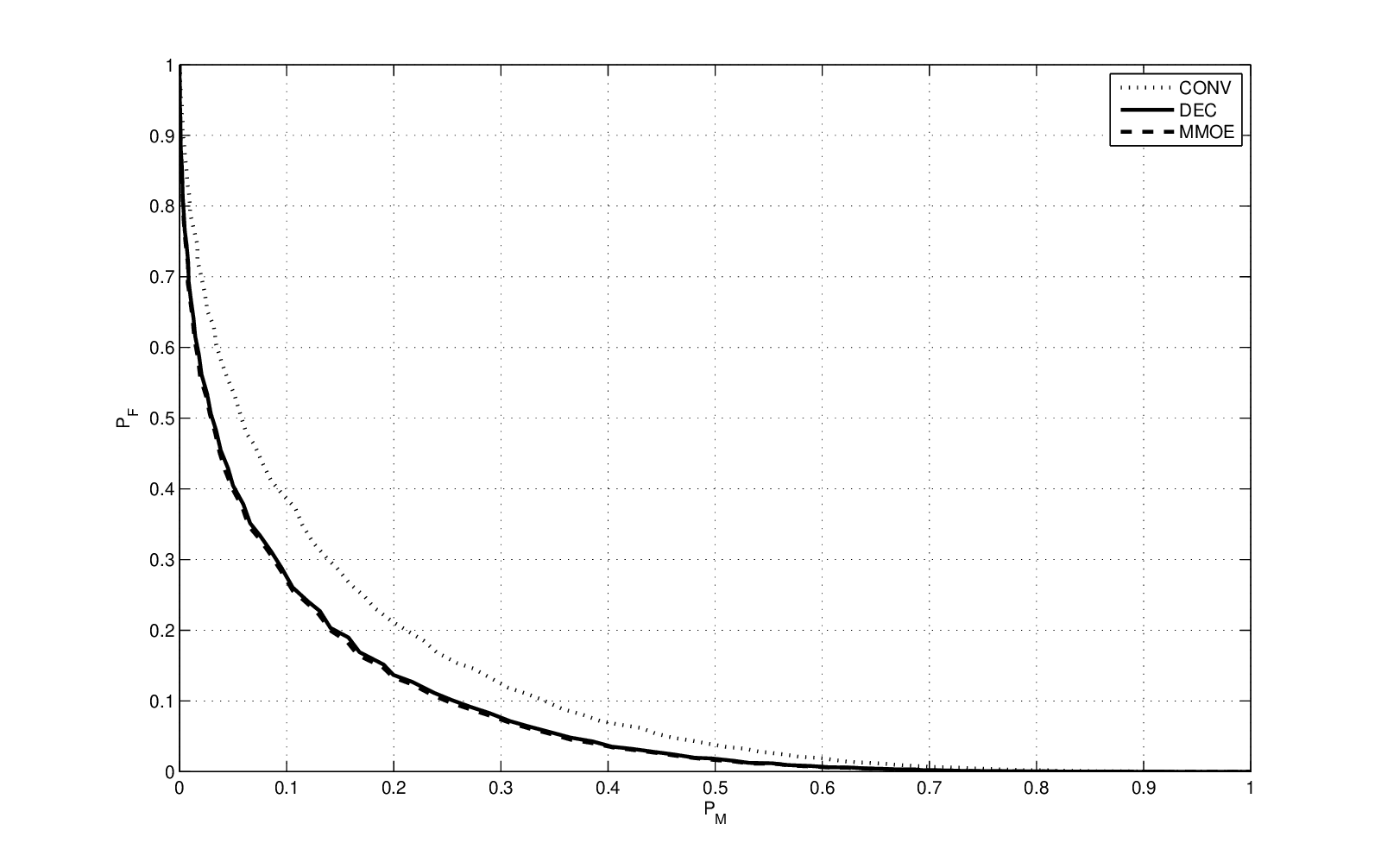}{\sl Performance of
various ND algorithms under perfect power control,
$2\sigma_1^2=2N_0=1$ (SNR$_1$=0 dB), $N=100$, fully-loaded
network.}{FIG1}
\insertfig{h}{0.8}{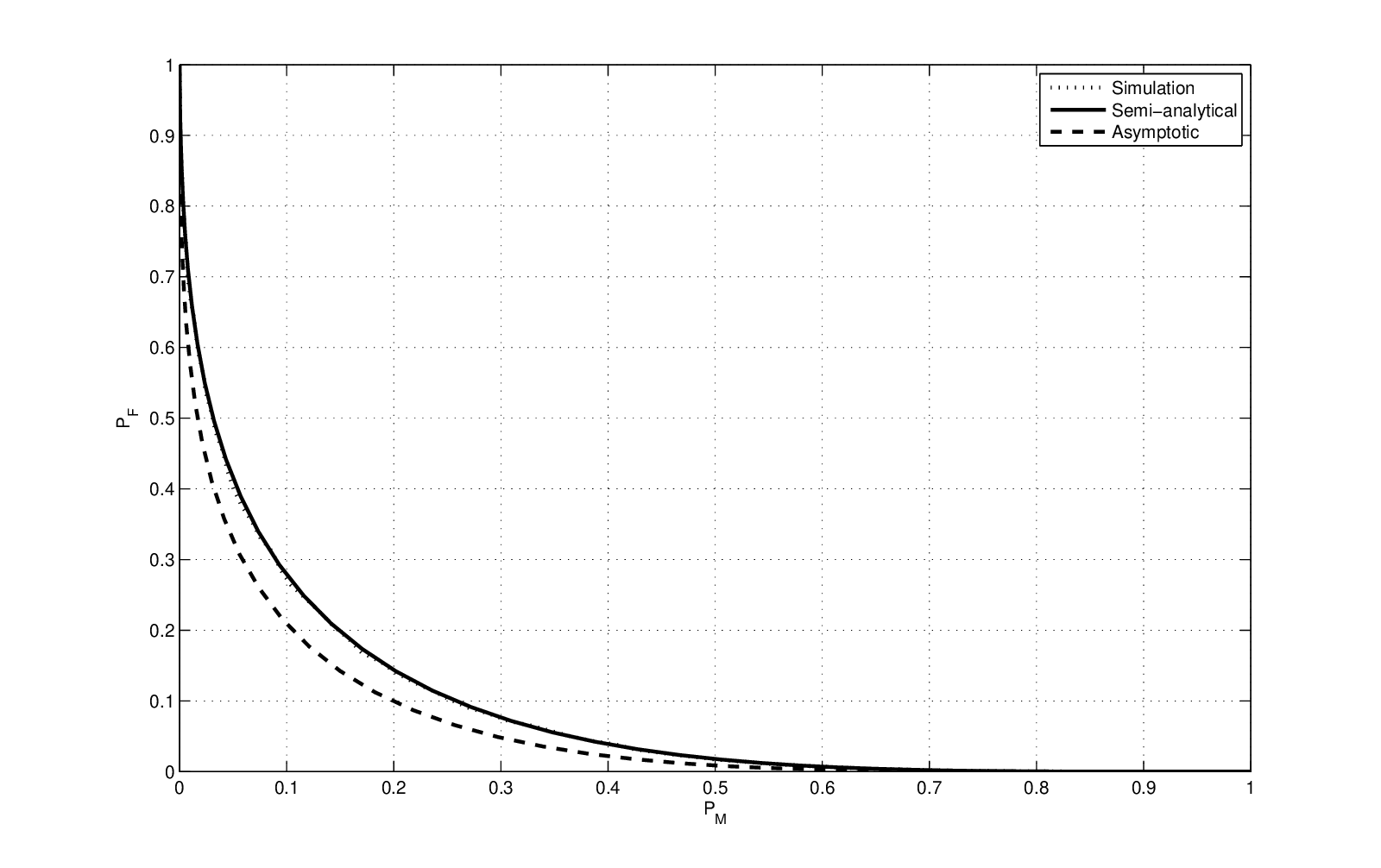}{\sl Performance of the CD for
$2\sigma_1^2=2N_0=1$ (SNR$_1$=0 dB), $N=100$}{FIG2a}
\insertfig{h}{0.8}{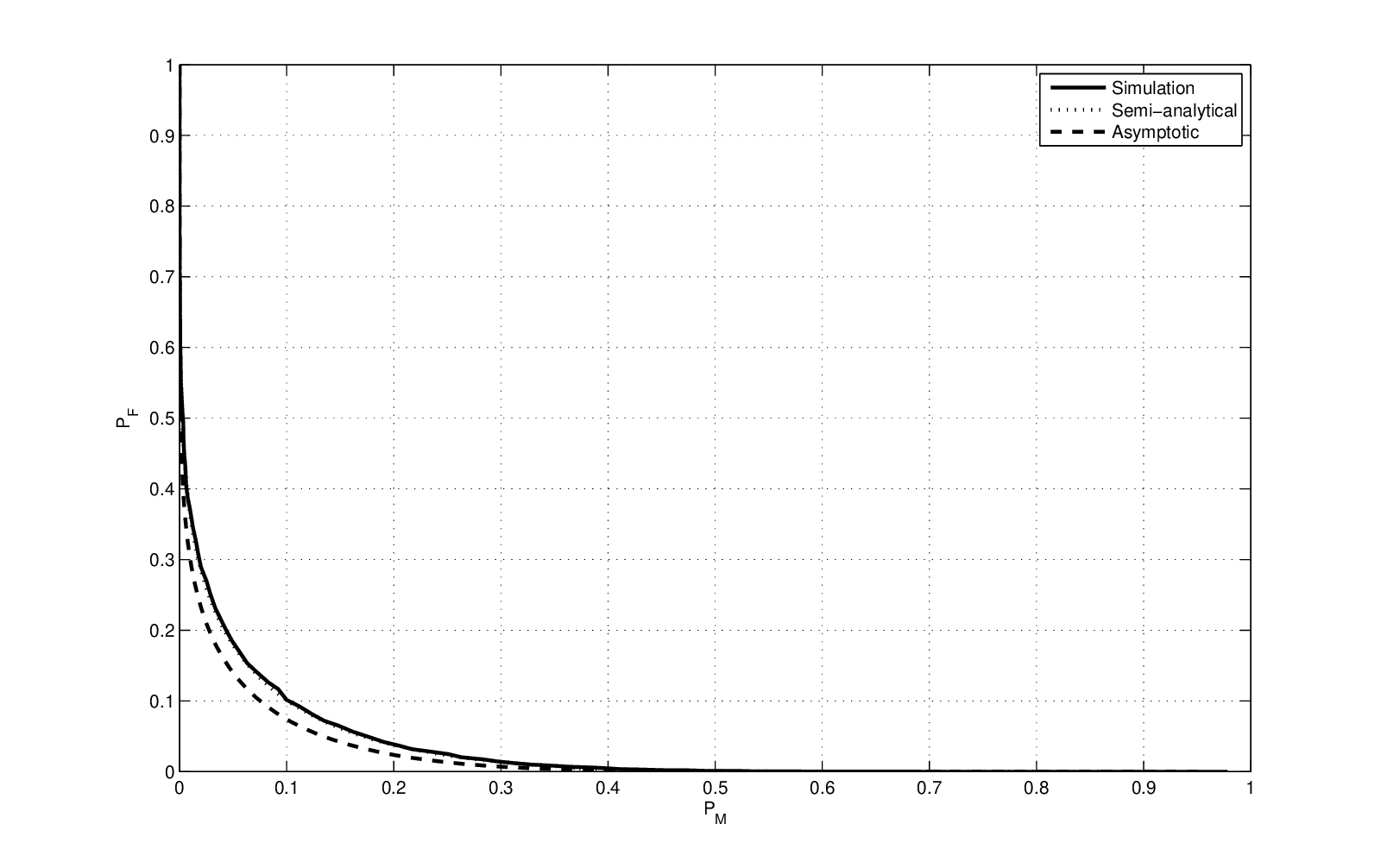}{\sl Performance of the CD for
$2\sigma_1^2=2N_0=1$ (SNR$_1$=0 dB), $N=300$.}{FIG2b}
\insertfig{h}{0.8}{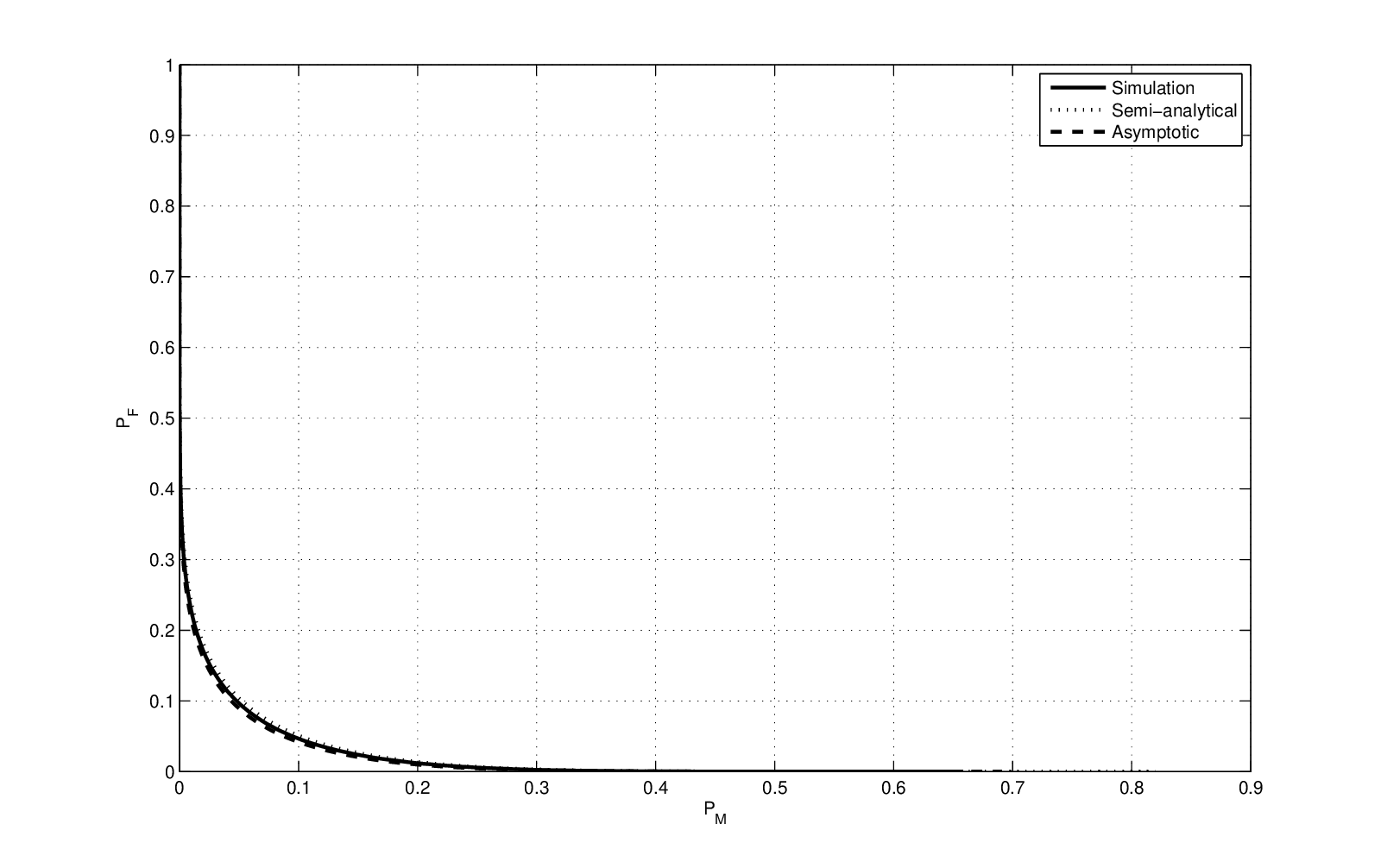}{\sl Performance of the CD for
$2\sigma_1^2=2N_0=1$ (SNR$_1$=0 dB), $N=500$}{FIG2c}
\insertfig{h}{0.8}{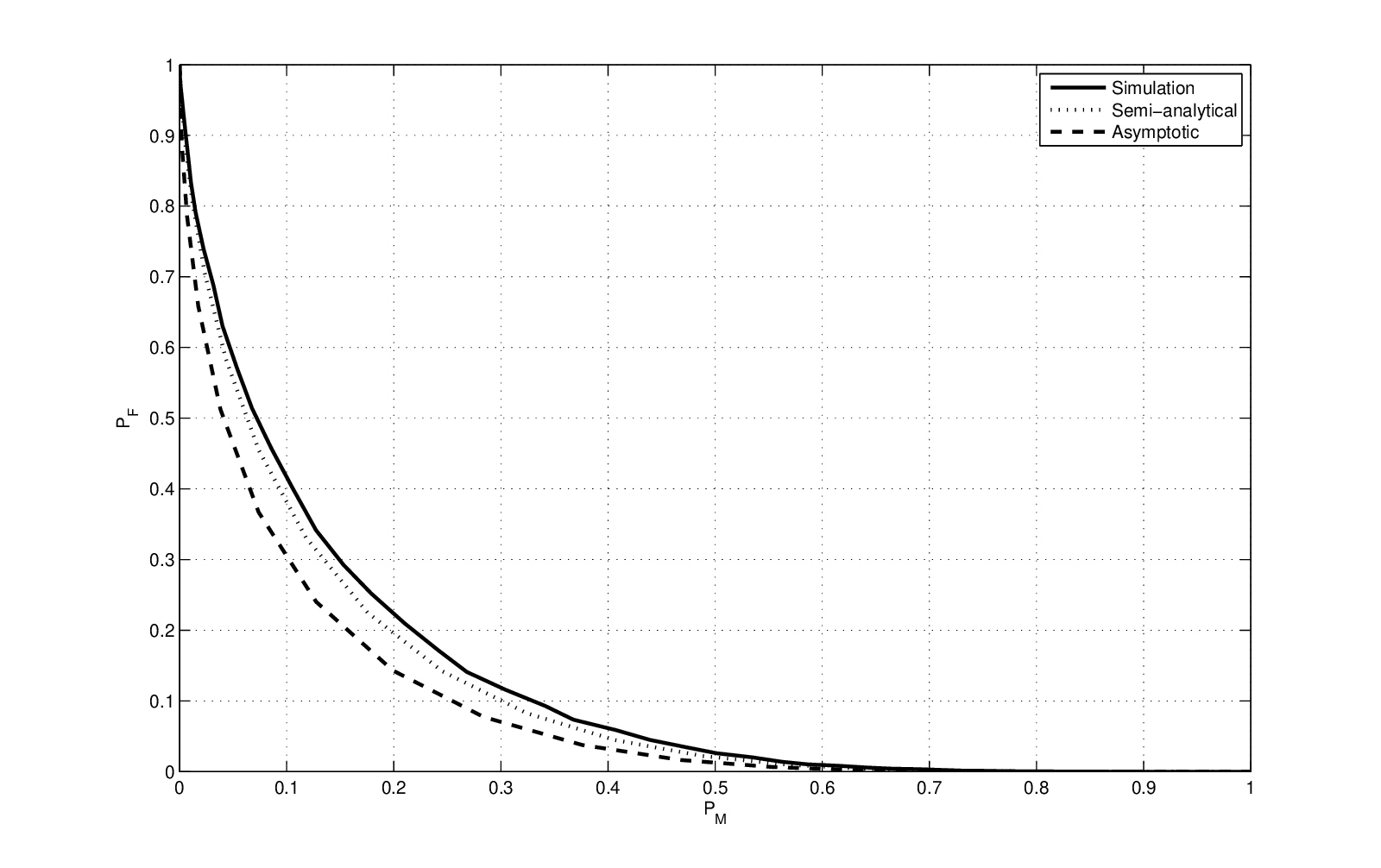}{\sl Performance of the ID for
$2\sigma_1^2=2N_0=1$ (SNR$_1$=0 dB), $N=100$.}{FIG3a}
\insertfig{h}{0.8}{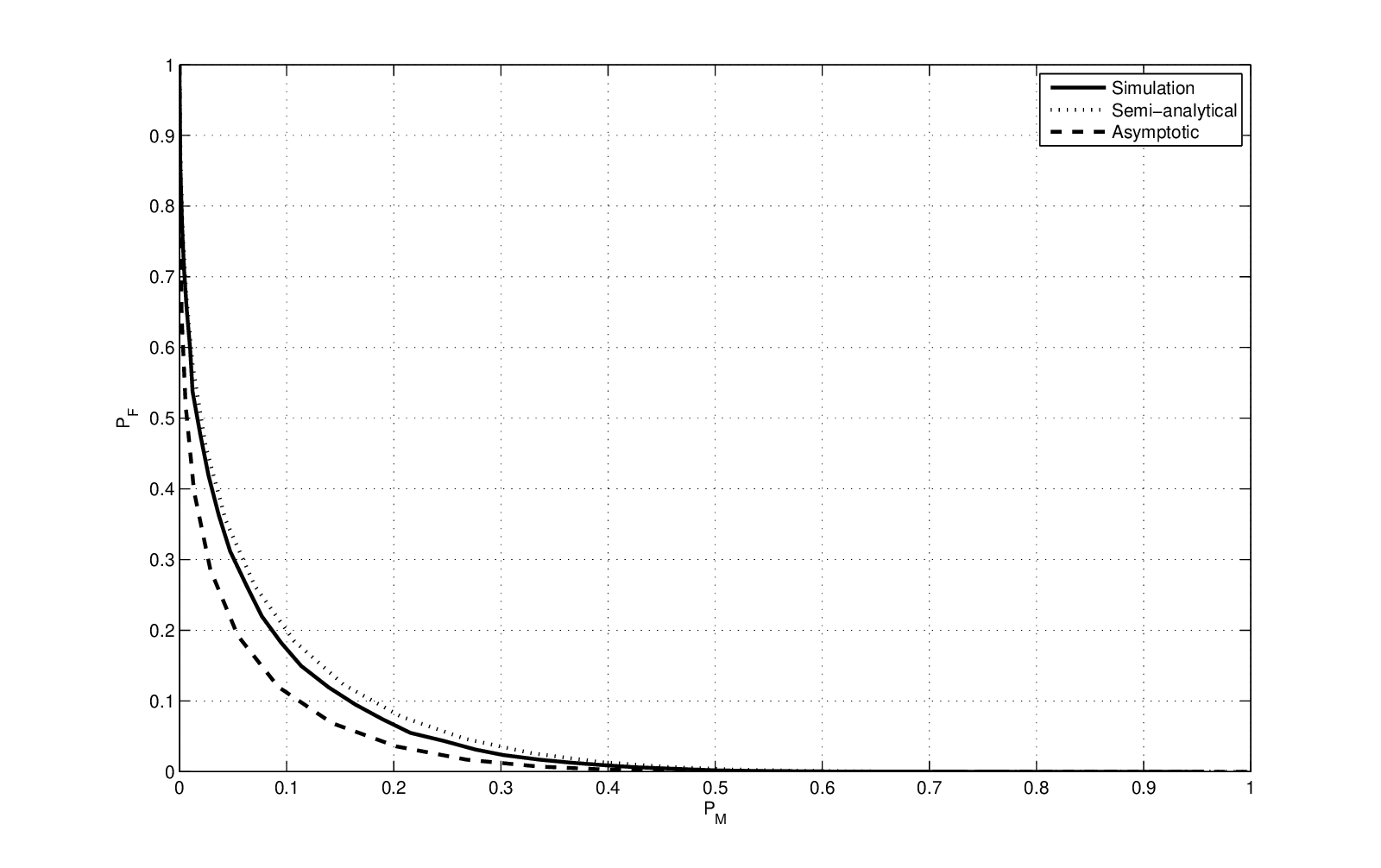}{\sl Performance of the ID for
$2\sigma_1^2=2N_0=1$ (SNR$_1$=0 dB), $N=300$}{FIG3b}
\insertfig{h}{0.8}{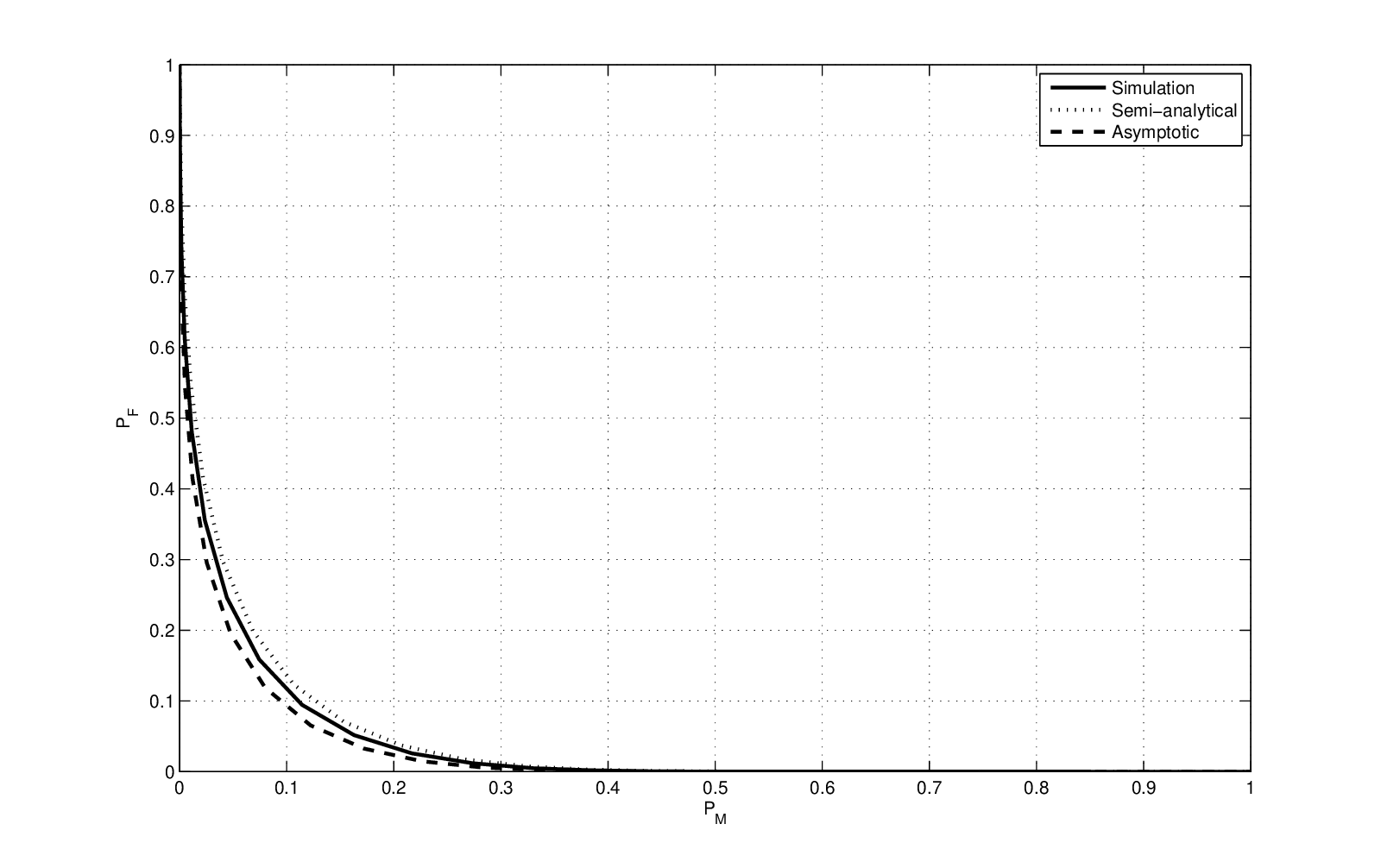}{\sl Performance of the ID for
$2\sigma_1^2=2N_0=1$ (SNR$_1$=0 dB), $N=500$.}{FIG3c}
\insertfig{h}{0.8}{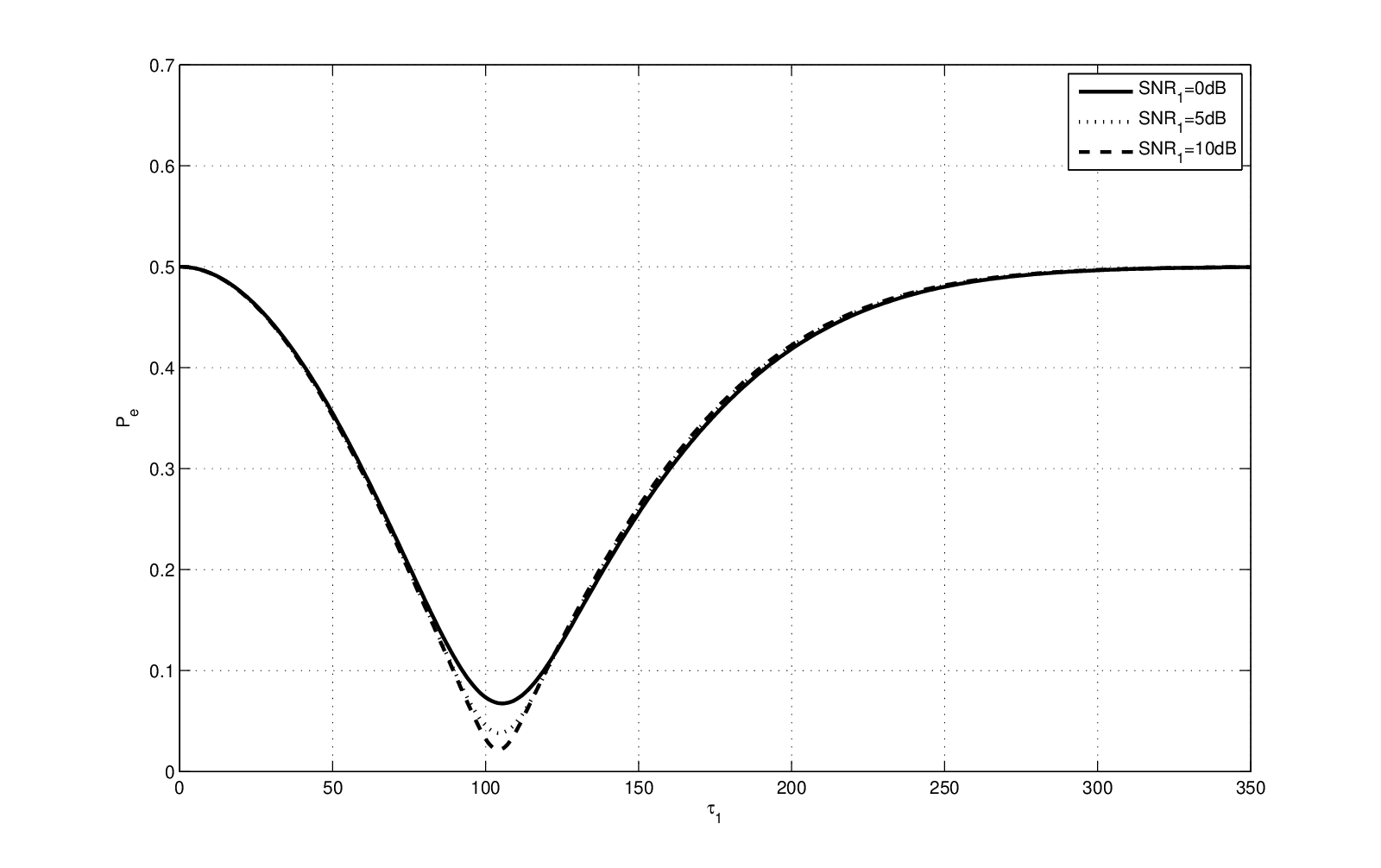}{\sl Error Probability versus the
detection threshold for a CD operating with $N=500$,
$\varepsilon=0.5$, $2\sigma_1^2=1$.}{FIG4}
\insertfig{h}{0.8}{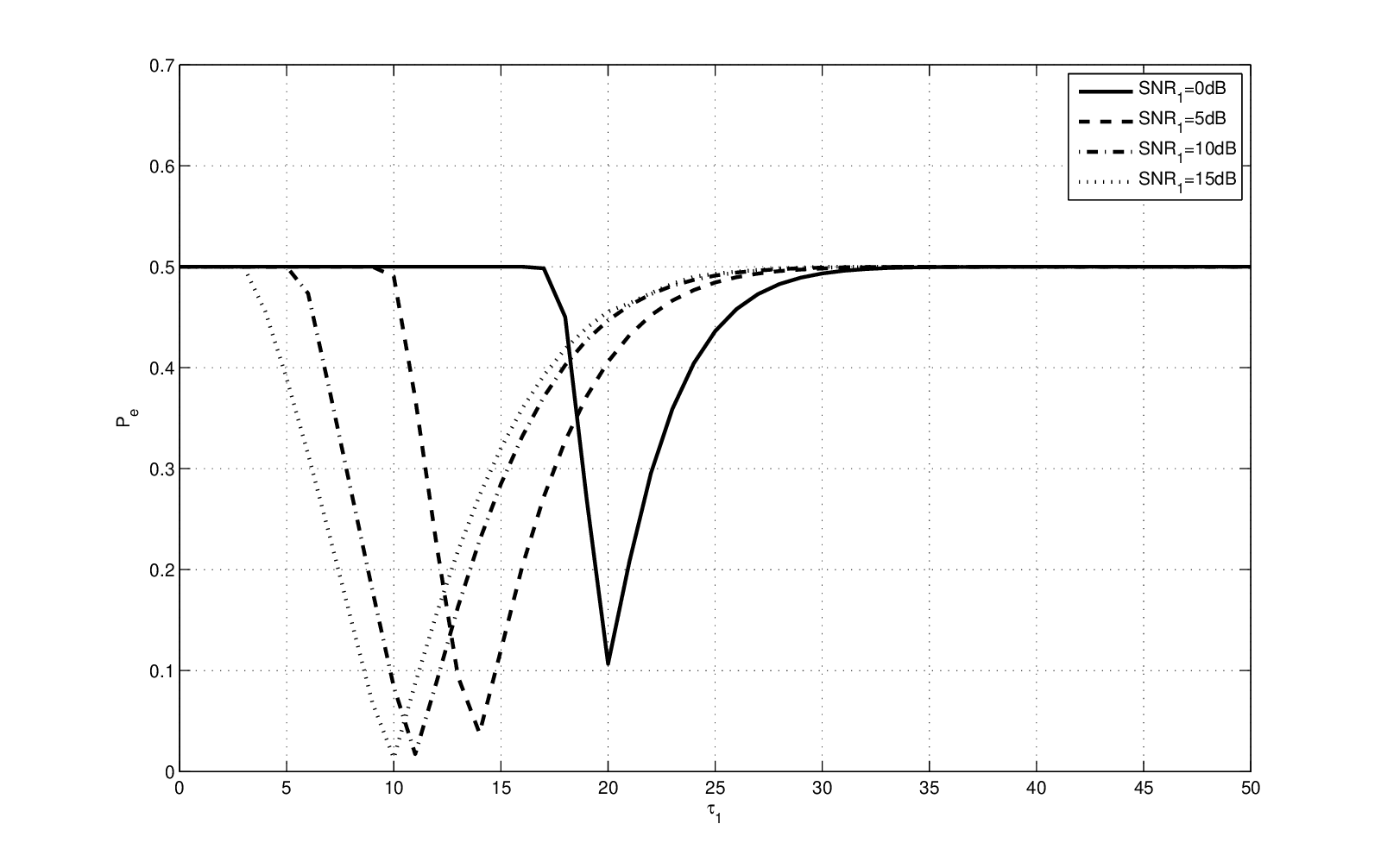}{\sl Error Probability versus the
detection threshold for an ID operating with $N=500$,
$\varepsilon=0.5$, $2\sigma_1^2=1$.}{FIG5}
\insertfig{h}{0.8}{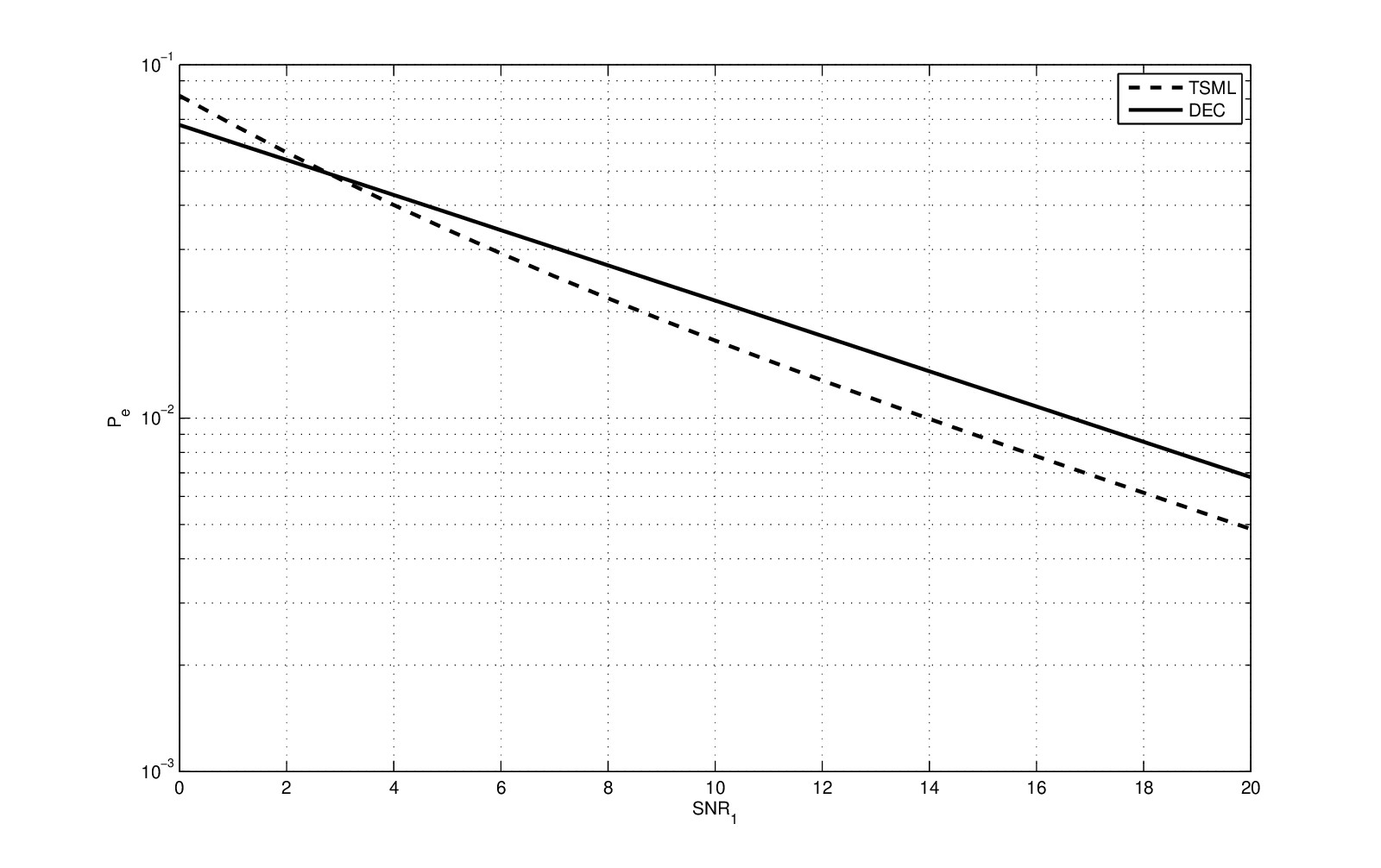}{\sl Global comparison between CD and ID,
$N=500$, $\varepsilon=0.5$, $2\sigma_1^2=1$.}{FIG6}
\end{document}